\documentclass{article}

\usepackage{moreverb,url}
\usepackage[affil-it]{authblk}
\usepackage[english]{babel}
\usepackage{blindtext}
\usepackage[colorlinks,bookmarksopen,bookmarksnumbered,citecolor=red,urlcolor=red]{hyperref}
\usepackage{graphicx}
\usepackage{multicol}
\usepackage[margin=1in]{geometry}

\renewcommand{\thefootnote}{\fnsymbol{footnote}}

\newcommand{\emath}{\ensuremath}
\newcommand{\nc}{\newcommand}
\newcommand{\ccdo}{CCD1}
\newcommand{\ccdt}{CCD2}

\newcommand{\lam}{\emath{\lambda}}

\nc{\PP}{\emath{\mathcal{P}}}
\nc{\mo}{Method I}
\nc{\mt}{Method II}
\nc{\Tbar}{\emath{\bar{T}(\lam)}}

\nc{\Smo}{\emath{S_{\mathrm{I}}(\lam)}}
\nc{\Smt}{\emath{S_{\mathrm{II}}(\lam)}}

\newcommand{\eref}[1]{Eq.~\ref{eq:#1}}
\newcommand{\erefs}[2]{Eqs.~\ref{eq:#1} and \ref{eq:#2}}

\newcommand{\fref}[1]{Figure~\ref{fig:#1}}
\newcommand{\frefs}[2]{Figures~\ref{fig:#1} and \ref{fig:#2}}
\nc{\cf}{{\sl{c.f.}}}
\nc{\eg}{{\sl{e.g.}}}

\begin{document}
\title{Calibration technique for suppressing residual etalon artifacts in slit-averaged Raman spectroscopy}
\author[1]{Christine Massie}
\author[2]{Keren Chen}
\author[1,2]{Andrew J Berger\thanks{\textbf{Corresponding author:}\\ Andrew J. Berger Ph.D\\Institue of Optics, University of Rochester, Rochester, NY, USA\\Email: andrew.berger@rochester.edu}}
\affil[1]{Department of Biomedical Engineering, University of Rochester, Rochester NY, USA}
\affil[2]{The Institute of Optics, University of Rochester, Rochester, NY, USA}
\date{}
\providecommand{\keywords}[1]
{
  \small	
  \textbf{\textit{Keywords---}} #1
}
\maketitle

\renewcommand{\thefootnote}{\arabic{ftest}}

\begin{abstract}
Back-illuminated charged-coupled device (BI-CCD) arrays increase quantum efficiency but also amplify etaloning, a multiplicative, wavelength-dependent fixed-pattern effect.  When spectral data from hundreds of BI-CCD rows are combined, the averaged spectrum will generally appear etalon-free.  This can mask substantial etaloning at the row level, even if the BI-CCD has been treated to suppress the effect.  This Note compares two methods of etalon correction, one with simple averaging and one with row-by-row calibration using a fluorescence standard.  Two BI-CCD arrays, both roughened by the supplier to reduce etaloning, were used to acquire Raman spectra of murine bone specimens.  For one array, etaloning was the dominant source of noise under the exposure conditions chosen, even for the averaged spectrum across all rows; near-infrared-excited Raman peaks were noticeably affected. In this case, row-by-row calibration improved the spectral quality of the average spectrum.  The other CCD's performance was shot-noise limited and therefore received no benefit from the extra calibration.  The different results highlight the importance of checking for and correcting row-level fixed pattern when measuring weak Raman signals in the presence of a large fluorescence background.
\end{abstract}



\section{Introduction}
Raman spectroscopy is a well-established inelastic scattering technique that produces a detailed molecular ``fingerprint" of vibrational modes present in a sample, including their relative abundances.  The technique can be performed upon optically thick specimens in a noncontact, backscattering geometry, requiring minimal sample preparation compared to Fourier transform infrared spectroscopy \cite{ir_sample}. In the biomedical sciences, Raman spectroscopy has been used to study various diseases and cancers in many different tissues \cite{invivoRaman,cancer,raman_biomed,raman_biomed2}. Such work typically uses near-infrared (NIR) excitation in the 780-840 nm range in order to increase the Raman/autofluorescence ratio.

Current slit-imaging Raman spectrometers commonly disperse light onto two-dimensional charged-coupled device (CCD) arrays. Back-illuminated (BI) CCDs have a higher quantum efficiency (QE) compared to front-illuminated CCDs because there is no obscuration from the front surface's readout wires~\cite{back_ill_ccd}.  Because the BI approach requires the  photosensitive region to be thinned to the scale of the optical absorption length, some photons pass through the whole region and reflect back, creating interference that manifests as a roughly sinusoidal oscillation in QE versus wavelength~\cite{Ma_etalon2,hu_etalon_bi}.
This QE variation, called the ``etalon effect'', affects all light, both the Raman scattering and the usually much stronger fluorescence background. The etalon effect becomes problematic in Raman spectroscopy when the chip-induced modulation in the fluorescence amplitude becomes comparable to the true Raman peaks' amplitudes. 
All else being equal, this problem is larger in the near-IR because the detector's absorption coefficient is lower, allowing more light to back-reflect.

To reduce the etalon effect in commercial practice, CCDs for imaging spectrometers are often surface-roughened in an attempt to decorrelate the QE values between adjacent pixels in both the horizontal (wavelength) and vertical (slit-height) directions.  This eliminates sinusoidal variations with larger periods but does not change the standard deviation of the QE fluctuations.  
The standard deviation can be reduced by suppressing back-reflection, either by increasing the depletion region thickness~\cite{spie_etalon3} or by depositing an wavelength-dependent anti-reflective (AR) coating across the CCD \cite{AR_coat}.  These options are rarely available commercially.

Rather than  modifying the CCD physically, one can reduce etalon effects computationally by comparing two or more spectra acquired in different grating positions~\cite{bell_subtract,ogrady_psd,auguie} or at different laser excitation wavelengths~\cite{cooper,kloz_fsrs}.  These approaches model the etalon contribution as additive and unchanged by shifting, such that subtraction will eliminate it but allow the Raman spectrum to be recovered. This is an approximate model; the etalon effect upon the spectrum is multiplicative, and shifting the wavelengths delivered to individual pixels (as in the grating shift approach) should alter the fixed pattern.

In applications where the slit's height dimension encodes no information of interest.   One such application,  single-point Raman spectroscopy of \textit{ex vivo} murine bone specimens, will be used  here as an example.  A laser illuminates a $\sim1$~mm  spot on the bone; the resulting Raman emission's etendue is enough to fill a slit and CCD array both a few mm in height.  The goal is to reduce the full 2D data matrix into a single  spectrum representative of the optically interrogated bone region.

This Note compares two methods of mitigating fixed pattern in such situations.  The first method simply sums the spectral rows (``vertical binning''). The other method employs row-by-row fixed-pattern correction, using a fluorescence standard as a pattern-free spectral reference.  This approach attempts to eliminate the fixed pattern effect rather than relying solely upon noise-averaging.  

\section{Methods}

\subsection{Instrument}
The Raman spectroscopy data were collected on a system described previously \cite{Shu,feng}. Briefly, a 830-nm laser delivered 150 mW to the sample in a 1 mm diameter spot. The scattered light was imaged onto a circular bundle of multimode fibers (100 micron core diameter, 0.27 NA). Light delivery and collection were performed in a non-contact manner. The proximal end of the bundle rearranged the fibers linearly at the entrance of a f/1.8 imaging spectrograph (Kaiser HoloSpec).  
Two deep-depletion, back-thinned, fringe-suppressed CCDs were tested.  ``\ccdo'' is an Andor Model DU 420-BX-DD, antireflection-coated for 750-1000 nm, purchased in 2001.  ``\ccdt" is an Andor Model DU-420-BEX2-DD, dual AR coating for UV-VIS-NIR broadband spectroscopy, purchased in 2018.  The vertical dimension of the detectors permitted spectral imaging of 40 fibers from the bundle.  

\subsection{Calibration}
The spectra were pre-processed for readout and dark noise and calibrated for wavelength and Raman shift using a neon lamp and Tylenol, respectively, as described previously \cite{jason_SORS}. 
Imaging distortions due to spectrograph optics and CCD alignment were corrected using a method described by Esmonde-White et al. \cite{white_image}. 
Laser excitation of a metal-ion-doped glass (Standard Reference Material 2241, National Institute of Standards and Technology [NIST]) provided spectrally smooth fluorescence without Raman peaks.  This enabled characterization of pixel-level fixed pattern effects, as described below. 
Each spectral row, ${n}$, of the resulting glass fluorescence data ${G_n(\lambda)}$ was divided by the NIST-supplied fluorescence emission spectrum ${F(\lam)}$ to produce a spectral throughput function ${T_n(\lambda)}$ in the form 
\begin{equation}
{T_n(\lambda)} = \frac{{G_n(\lambda)}}{{F(\lambda)}}.
\label{eq:throughput}
\end{equation}
This forced the rendered glass spectra from each row to have the same spectral lineshape, although overall amplitudes differed due to the illumination and collection geometry.   

Using this terminology, \mo\ (fixed pattern suppressed solely by averaging) produces the output spectrum
\begin{equation}
\Smo = \sum_n \PP \big[ S_n(\lam) \big / \Tbar],
\label{eq:methodOne}
\end{equation}
where $\PP$ denotes post-processing [subtraction of an estimated fluorescence background via continuous wavelet transform~\cite{cwt} followed by Savitzky-Golay smoothing over a 3 pixel window \cite{golay_filter} to match the spectral resolution of 6 cm$^{-1}$] and \Tbar\ denotes the average of the $T_n(\lam)$ spectral throughput functions.  
Similarly, \mt\ (fixed pattern corrected row by row) produces the output spectrum
\begin{equation}
\Smt = \sum_n \PP \Big[ S_n(\lam)/ T_n(\lam) \Big].
\label{eq:methodTwo}
\end{equation}
Note that in \eref{methodTwo} each row spectrum is corrected for its particular fixed pattern, unlike in \eref{methodOne} where all rows are corrected using the same function $\bar{T}(\lam)$.   
As an optional intermediate step, signals from individual optical fibers were sometimes calculated.  For qualitative analysis of fixed pattern variation across the CCD array, ten adjacent rows of an unprocessed spectral image were simply averaged at each of three locations (see \fref{2}). For quantitative estimates of each fiber's spectrum, a Gaussian-Lorentzian decomposition described by Dooley et al.\ \cite{dooley}) corrected for spatial overlap between adjacent fibers' images (see \frefs{3}{4}).

{\subsection{Specimens}
The experimental data presented here were acquired in two previous studies \cite{Shu,massie}. Briefly, a tibia \cite{Shu} and femur \cite{massie} were excised from C57BL/6J mice between the age of 4-24 weeks. The exposed bone specimens were stored in phosphate buffered saline and frozen before performing Raman spectroscopy.}

\section{Results}

\fref{1} compares unprocessed fluorescence measurements of the same calibration glass using the two different CCD arrays.
The summed spectra over all rows (panels A and B) exhibit smooth lineshapes without visible etalon effect artifacts for either array.  Ripples characteristic of the etalon effect are evident in three adjacent row spectra from \ccdt\ (panel D); such ripples are absent for \ccdo.  High-pass-filtered spectra from  two adjacent rows of \ccdt\ (panel F) share a common ripple effect from the etalon effect but have row-specific variations at the pixel level due to both fixed pattern and shot noise.  Oscillations are not seen for pixels 1-400 because the spatial period is less than two pixels.  Analogous spectra from \ccdo\ exhibit lower amplitude variations with no correlation between rows;  successive exposures of the same row also indicated that fixed pattern was not the dominant noise source (data not shown). \\ 

\begin{figure}[h]
    \centering
    \includegraphics[width = 2in]{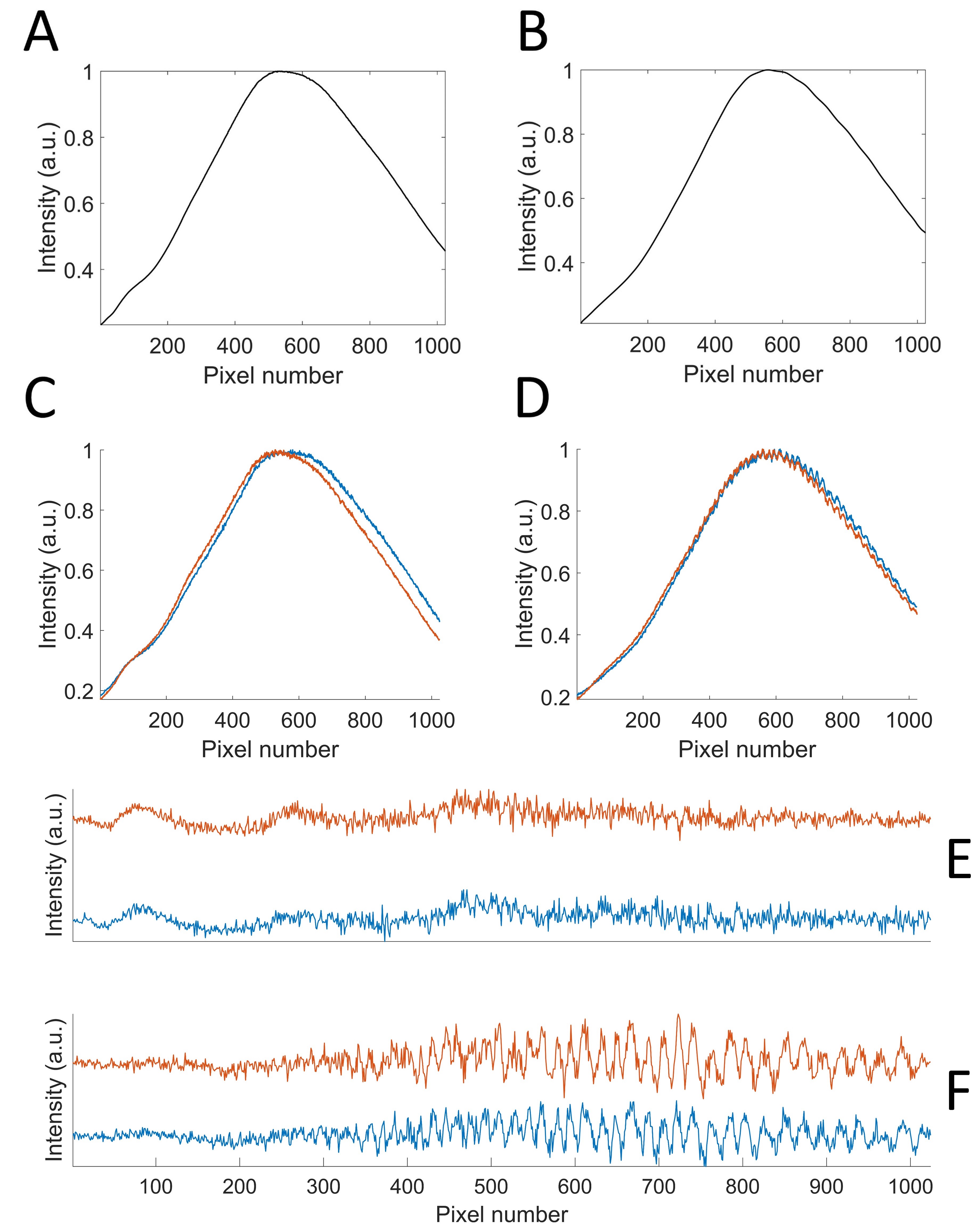}
   \caption{Comparisons of unprocessed metal-ion-doped glass data measured with \ccdo\ (A,C,E) and \ccdt\ (B,D,F) normalized to their maxima.  Wavelength increases linearly with pixel number.  (A,B) Sum of all rows of spectral data collected on \ccdo\ (A) and \ccdt\ (B).  Spectra look similar with no visible fixed pattern. (C,D) Neighboring individual rows from \ccdo\ (C) and \ccdt\ (D) show a more distinct etalon effect present with data collected on \ccdt.
(E,F) High-pass filtered versions of two spectra from each of C,D. Traces are offset for clarity; all amplitudes of fluctuation are plotted on the same scale for direct comparison.}
\label{fig:1}
\end{figure}

\begin{figure}
\centering
\includegraphics[width = 2in]{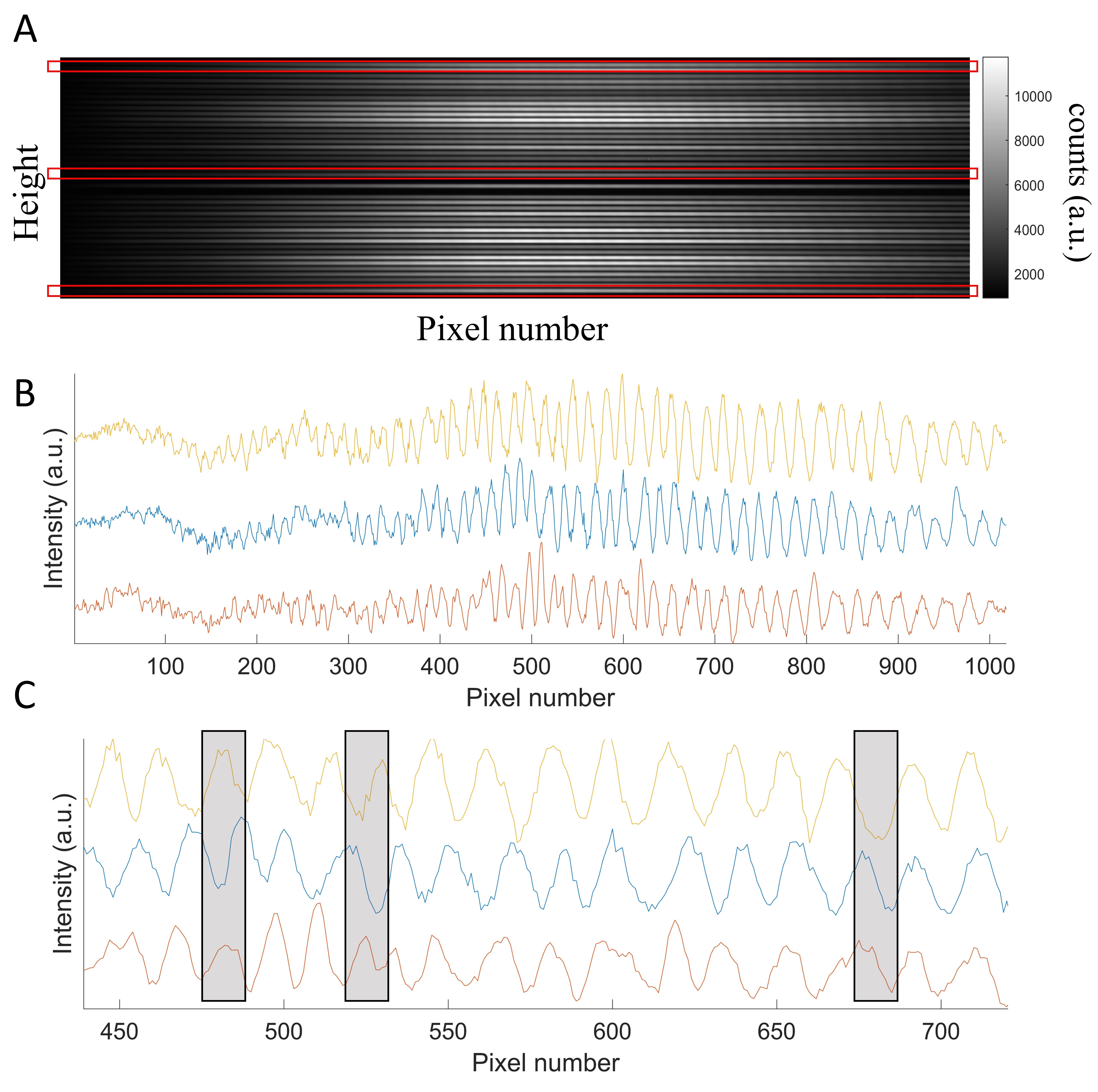}
\caption{(A) Image collected from metal-ion-doped glass on CCD 2. Multiple adjacent rows are summed to create a spectrum from an optical fiber. The red boxes outline non-neighboring fibers that were used to characterize fixed pattern. (B) Residuals after subtracting a smoothed fit to three non-neighboring fiber spectra showing the fixed pattern is not uniform throughout the height of the CCD. (C) Zoomed in version of residuals shown in B with grey boxes highlighting a few locations showing the fixed pattern is not uniform. }
\label{fig:2}
\end{figure}

\fref{2}A shows the uncalibrated \ccdt\ spectral image of glass fluorescence from which Figures 1B/D/F were calculated, with each horizontal stripe corresponding to one optical fiber's output.  Data from optical fibers at three different array heights (marked by red boxes) were vertically binned as noted previously and then high-pass filtered, producing the spectra shown in Fig.\ref{fig:2}B.  Oscillatory patterns similar to \fref{1}F  are seen; the spectra are smoother because multiple rows are summed, averaging individual row's pixel-level variations.  Grey bars in \fref{2}C emphasize that  etalon oscillations are dephased between fibers at different array heights.  This explains why etalon effects wash out in a simple sum over all fibers' contributions, as was shown in \ref{fig:1}B. \\

Methods I and II  were applied to Raman spectroscopy of an  \textit{ex vivo} murine bone specimen collected using \ccdt. \fref{3}A-B shows total spectra (Raman plus fluorescence) from individual fibers; panels C-D are the corresponding background-subtracted (``Raman only'') versions.  
As expected, the \mo\ fiber spectra exhibit much greater oscillatory patterns, due to the lack of local fixed pattern correction. 
Adding all fibers' contributions together via the two methods produced the background-corrected spectra E and F in \fref{3}; these are \Smo\ and \Smt\ from \erefs{methodOne}{methodTwo}, respectively.  The main features of \Smt\ are those expected of a typical bone spectrum \cite{morris_contribution}. \Smo, by contrast, contains a noticeable oscillatory component throughout the spectrum.  Plot H provides a magnified view of this additional noise in \Smo\ distorting the lower-amplitude Raman peaks from the bone.  \\

\begin{figure}[h]
\centering
\includegraphics[width = 2in]{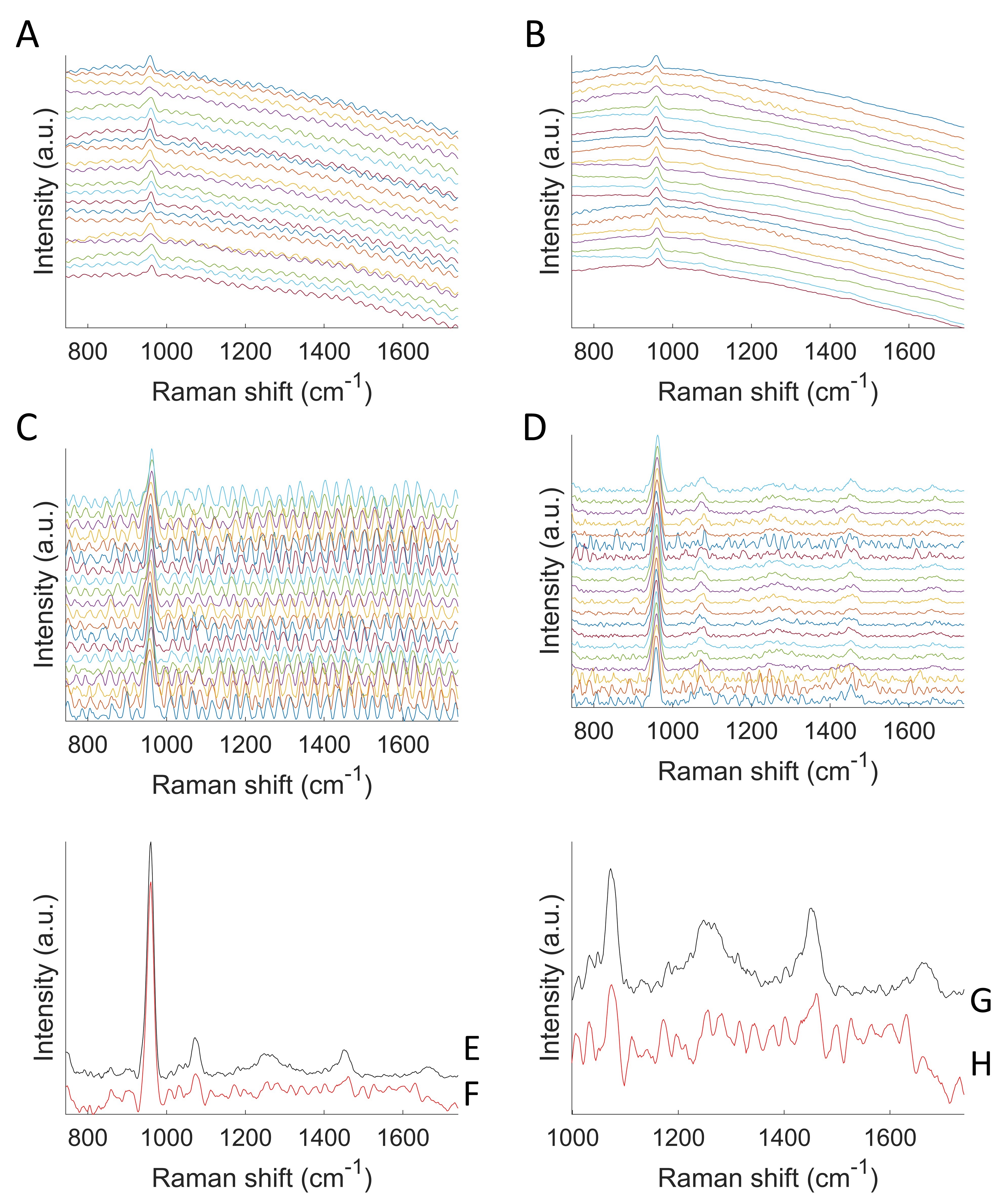}
\caption{Data collected with \ccdt and calibrated using \mo\ (A,C,F,H) and \mt\ (B,D,E,G). (A-B) Full spectra (Raman plus fluorescence) from individual fibers.  The etalon effect is seen for \mo\ but not for \mt.  (C-D) Polynomial subtracted and normalized (to mean absolute deviation) spectra using (C) \mo\ and (D) \mt. (E-F) Final output spectra (E) \Smt\ and (F) \Smo.  The \Smo\ spectrum contains a residual amount of the etalon effect.  (G-H) Zoomed in region demonstrating the distortion of minor Raman peaks from bone in \Smo\ due to uncorrected fixed pattern.}
\label{fig:3}
\end{figure}

\fref{4} shows analogous bone spectra measured from \ccdo. 
In this instance the \Smt\ spectrum provided by \mt\ provided no obvious improvement over \Smo.

\begin{figure}[h]
    \centering
    \includegraphics[width = 2in]{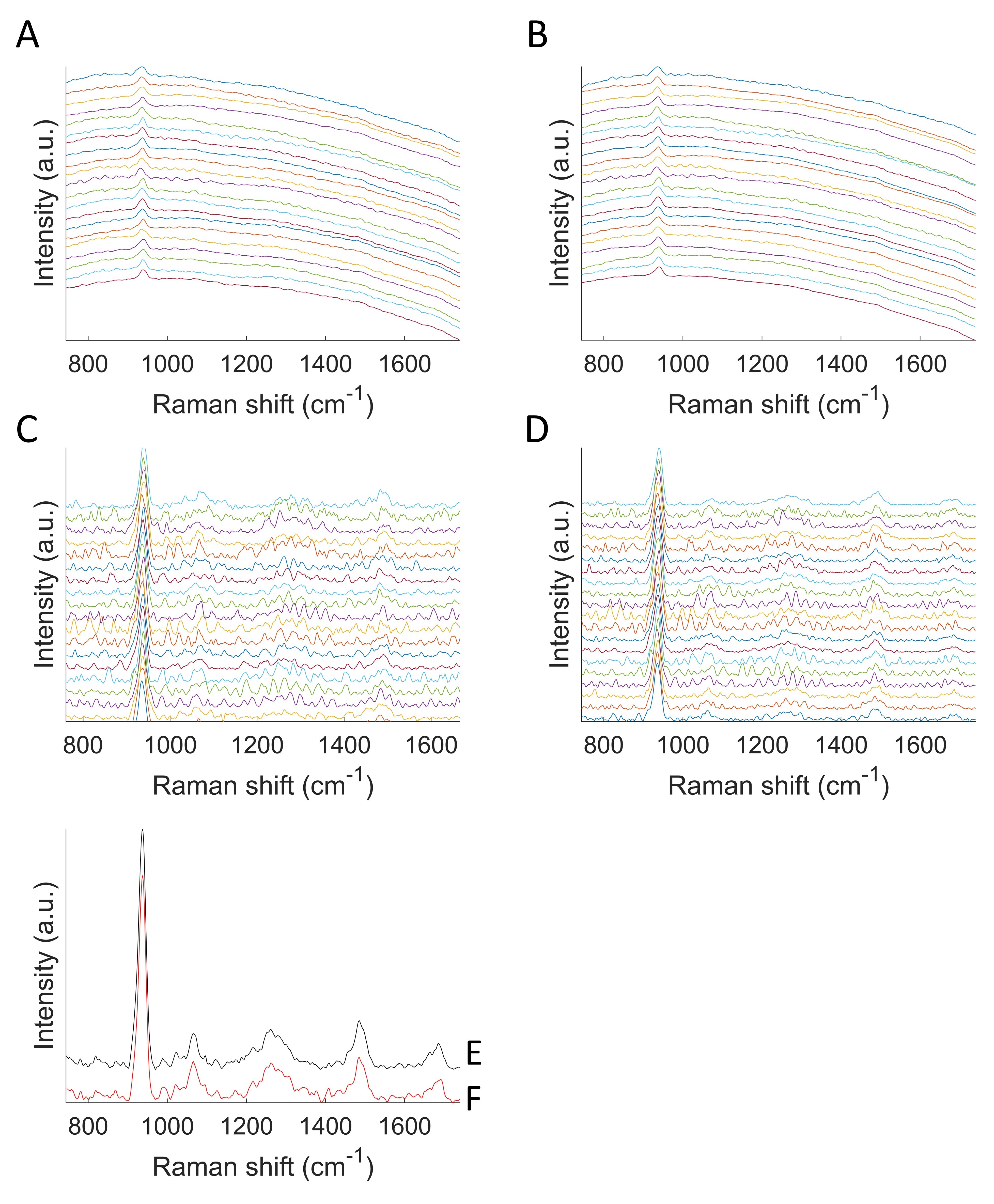}
    \caption{Data collected with \ccdo\ and calibrated using \mo\ (A,C,F) and \mt\ (B,D,E). (A-B) Calibrated total-signal spectra (Raman plus fluorescence) from individual fibers using (A) \mo\ and (B) \mt.   
    (C-D) Same spectra, polynomial-subtracted and normalized. (E-F) Final output spectra (E) \Smt\ and (F) \Smo, with neither spectrum exhibiting the oscillatory component present in \fref{3}F.}
    \label{fig:4}
\end{figure}

\section{Discussion}
Back-illuminated CCD chips for spectroscopy provide higher overall QE with the downside of greater QE variation due to etalon effects.  When spectroscopy hardware or software is set to bin over many spectral rows, the residual effect of the etalon effect can be impossible to perceive in a plot, as in the \mo\ spectrum of \fref{1}A.  This could lead a user to conclude that two CCDs have equivalent levels of fringe suppression.    

In the case of Raman peaks riding on a fluorescence background, this metric for equivalency can be insufficient.  Two CCDs, marketed similarly by a major supplier, exhibited qualitatively different levels of the etalon effect.  Notably, the one manufactured more recently (\ccdt) had the larger levels. In this instance, the fixed pattern from the etalon effect was readily visible in single rows (\cf\ \fref{1}) and contributed significant noise to the vertically binned spectrum \Smo.  Row-by-row calibration of the fixed pattern was required to render lower-amplitude Raman peaks more clearly (\cf\ \fref{3}).  This indicates the importance of correcting fixed pattern when measuring Raman features in the presence of large fluorescence backgrounds, even when binning over the height of a slit. 

On the other hand, \ccdo\ did not exhibit fixed pattern at the row level for the same experimental settings.  There was therefore no expected benefit to using row-level calibration (\cf\ \fref{4}).  In fact, if the noise amplitude had been a greater fraction of the total signal, \mt\ should have been noticeably detrimental, as it imprinted shot noise from the glass measurements onto the bone spectra. In this particular study, the signal to noise ratio was high enough that the two methods produced approximately equivalent results.

In this work we used a translucent glass calibration sample and an opaque, highly scattering bone specimen.  The angular distributions of light emission from these two targets were unequal, leading to different relative amounts of light collected by different optical fibers in the two cases.  This in turn causes incorrect weighting of the fiber contributions to the final spectrum, but does not affect the correction of fixed pattern.

\newpage
\bibliographystyle{ieeetr}
\bibliography{mybibliography}

\end{document}